\documentclass[aps,prl,twocolumn,groupedaddress,showpacs,amssymb,amsmath]{revtex4}
\usepackage{graphicx}

\usepackage{color}
\definecolor{red}{rgb}{1,0,0}
\definecolor{blu}{rgb}{0,0,1}
\definecolor{mag}{cmyk}{0,1,0,0.5}
\newcommand{\gtap}{\mbox{$^{_{\textstyle >}}\!\!\!\!\!_{_{\textstyle \sim}}$}}
\newcommand{\ltap}{\mbox{$^{_{\textstyle <}}\!\!\!\!\!_{_{\textstyle \sim}}$}}

\begin{document}



\title{Energetics of midvelocity emissions
in peripheral heavy ion collisions at Fermi energies}


\author{A.~Mangiarotti},
\author{P.R.~Maurenzig},
\author{A.~Olmi},
\author{S.~Piantelli},
\author{L.~Bardelli},
\author{A.~Bartoli},
\author{M.~Bini},
\author{G.~Casini},
\author{C.~Coppi},
\author{A.~Gobbi}
\thanks{retired from GSI, Darmstadt}
\author{G.~Pasquali},
\author{G.~Poggi},
\author{A.A.~Stefanini},
\author{N.~Taccetti}
\author{E. Vanzi}

\affiliation{Sezione INFN and Universit\`a di Firenze, 
            Via G. Sansone 1, I-50019 Sesto Fiorentino, Italy}


\date{\today}

\begin{abstract}
Peripheral and semi-peripheral collisions have been studied in the
system $^{93}$Nb+$^{93}$Nb at 38 AMeV. 
The evaporative and midvelocity components of the light charged
particle and intermediate mass fragment emissions have been
carefully disentangled. 
In this way it was possible to obtain the average amount not only 
of charge and mass, but also of energy, pertaining to the midvelocity
emission, as a function of an impact parameter estimator.
This emission
has a very important role in the overall 
balance of the reaction, as it accounts for a large fraction of the
emitted mass and for more than half of the dissipated energy.
As such, it may give precious clues on the microscopic mechanism of
energy transport
from the interaction zone toward the target and 
projectile remnants.

\end{abstract}


\maketitle

In peripheral to mid-central collisions of heavy ions at 
\textit{Fermi energies} (E$_\mathrm{\,beam}$= 30--50 AMeV) 
an intense emission of light charged particles (LCP) and especially of
intermediate mass fragments (IMF) at velocities between those of the
projectile- and target-like fragments (PLF and TLF) is observed 
\cite{Montoya94,Toke95,Dempsey96,Lukasik97,Plagnol99,Davin02}.
These so-called ``neck-emissions'' or ``midvelocity emissions'' 
may be seen as an intermediate stage between the fast 3-body 
processes found at lower energies \cite{Casini93,Bocage00} and
the explosion of the ``participant zone'' at much higher energies.

At low beam energies ($\ltap$15 AMeV), the exchange of nucleons
between PLF and TLF, whose mean fields merge for a prolonged time,
plays a dominant role in the energy dissipation process,
but it is very difficult to have experimental access to
probes directly connected to the mechanism of matter exchange.
At high energies ($\gtap$100 AMeV), because of the small de~Broglie
wavelength of the nucleons and the reduced Pauli blocking,
the energy dissipation mechanism is dominated by direct nucleon-nucleon
collisions in the overlap between projectile and target, and
the short interaction time does not allow a significant heat transport
to the non-overlapping zones.
At Fermi energies the energy dissipation is already significantly
localized in the midvelocity region -- as quantitatively demonstrated
in this Letter --, while the interaction time may still be long enough
to allow the transfer of a sizable amount of energy to the surrounding
nuclear matter thus resulting in an excited PLF and TLF 
(henceforth indicated with PLF$^\star$ and TLF$^\star$).

To clarify the mechanism leading to midvelocity emissions, 
an important aspect is represented by the amount of energy 
involved in these emissions 
{ \cite{Larochelle99}}, 
as compared with the excitation energy left in the PLF$^*$ or TLF$^*$.
This may yield clues about the transition between the low-energy 
dissipative collisions (where the energy removed from the relative 
motion is totally converted into excitation of PLF$^\star$ and
TLF$^\star$) and the high-energy participant-spectator regime. 
We present here, for the first time, a direct simultaneous
determination of the energy involved in the midvelocity and 
in the evaporative emissions. 

The results of this Letter refer to the collision 
$^{93}$Nb+$^{93}$Nb at 38 AMeV, as a part of a systematic study 
\cite{Piantelli01,Piantelli02,Mangiarotti03,Piantelli}
of heavy ion collisions performed at the Superconducting Cyclotron of
the Laboratori Nazionali del Sud of INFN in Catania, 
with the {\sc Fiasco} setup \cite{Bini03}.
The apparatus consists of gas detectors (measuring heavy products with
E$\gtap$0.1 AMeV, in $\approx 70\%$ of the forward hemisphere), 
$\Delta E-E$ Silicon telescopes (determining charge and mass of PLFs
below the grazing angle) and phoswich telescopes (for isotopic
identification of LCPs and elemental identification of IMFs with
$Z\geq 3$, in about $30\%$ of the forward hemisphere).

This Letter is focused on two-body events, by far prevailing in
(semi-)peripheral collisions, where two and only two large 
reaction remnants are detected by the gas counters 
(which are fully efficient for heavy fragments with Z$\gtap$10--14).
This allows to have as clean as possible kinematics, profiting in the
best way of {\sc Fiasco}'s ability to detect also the TLF, with the
aim of studying the energy associated to midvelocity emissions in a
situation where their release represents the dominant process. 

The setup measures the secondary quantities of PLF and TLF
after the sequential evaporation, while primary quantities 
of the excited PLF$^\star$ and TLF$^\star$ are estimated from the
measured velocities with the kinematic coincidence
technique~\cite{Casini89}. 
As an indicator of the centrality of the collision
we have chosen a kinematic variable, defined as
  TKEL  = $E_\mathrm{\,in}^\mathrm{\,c.m.}\, -\,
   \tfrac{1}{2}  \,  \Tilde{\mu}\,
   { v_{\mathrm{rel}}^{\,2}}      
  $,
where $E_\mathrm{\,in}^\mathrm{\,c.m.}$ is the center-of-mass (c.m.)
energy in the entrance channel, $\Tilde{\mu}$ is the
reduced mass calculated with the masses of the kinematic
reconstruction (forced to add up to the total mass of the system and
thus overestimated) and  ${ v_{\mathrm{rel}}^{\,2}}$ is
the reconstructed primary relative velocity 
between PLF$^\star$ and TLF$^\star$.
While at low incident energies (where reactions are strictly binary),
TKEL truly represents the Total Kinetic Energy Loss of the collision, 
it is important to emphasize that in this work, where a sizable
midvelocity emission is present, TKEL is used just as an order
parameter for classifying the events in bins of different impact
parameter.  
Indeed, the kinematic analysis of events generated by a 
Quantum Molecular Dynamics (QMD) code \cite{Lukasik93} shows a strict
correlation between TKEL and impact parameter \cite{Piantelli01}.

\begin{figure}
\includegraphics[width=70mm,bb=10 20 350 210]{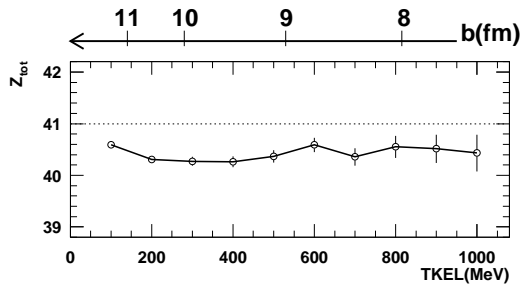}
\caption{\label{fig:ztot}
  Average total charge of forward-going products.
  The dotted line indicates the charge of a Nb projectile, while
  the arrow on top shows the impact parameter scale estimated 
  from QMD calculations \cite{Lukasik93}.
}
\end{figure}

The average multiplicities of charged particles,  
$\langle\mathcal{M}_{C_i}\rangle$
(henceforth we use the brackets $\langle \; \rangle$
to indicate averages over events in a given bin of TKEL),
were obtained from the distributions of the experimental yields of 
p, d, t, He and IMFs in the ($v_{\,\perp}$,$v_{\,\parallel}$) plane,
after correcting for the finite geometrical coverage of the phoswiches.
Here $v_{\,\parallel}$ and $v_{\,\perp}$ are the velocity components
parallel and perpendicular, respectively, to the asymptotic
PLF$^*$-TLF$^*$ separation axis in the c.m. system;
for the TKEL range addressed in this work, the c.m. separation axis 
lies within about 10$^\circ$ from the beam axis.
The advantage of using a symmetric collision is that the forward-going
particles (those with $v_{\,\parallel}\geq 0$) must have the same
average characteristics as the backward-going ones.  
Therefore, in this paper all multiplicities refer only to forward
emitted particles, for which the setup has a much better solid angle
coverage and threshold effects do not play any role 
(all particles having lab-energies larger than $\approx$~9.5 AMeV).
In a symmetric system, adding up the charges of forward emitted LCPs
and IMFs to the charge of the PLF should reproduce, on average, the
projectile charge. 
The deficit of about half a charge unit shown in
Fig.~\ref{fig:ztot} was corrected by slightly rescaling the
contribution of all charged reaction products in each TKEL bin.

Another advantage of a symmetric collision is that the contribution of
the free neutrons (unmeasured) can be estimated from mass conservation.
In fact the average multiplicity of all the undetected forward-emitted
free neutrons is given by 
$ \langle\mathcal{M}_{n}\rangle =  A_\mathrm{proj} 
    - \sideset{}{_i} \sum A_{C_i}  \langle\mathcal{M}_{C_i}\rangle 
    - \langle A_\mathrm{sec}\rangle $,
where $\langle A_\mathrm{sec}\rangle$ is the average secondary mass of
the PLF (measured with the Silicon telescopes) and $A_{C_i}$ are the
mass numbers of the different charged particle species.
In doing so, the common assumption $A=2\,Z$ was used for estimating
the masses of the IMFs, which were not isotopically resolved. 
This seems reasonable, since 
experimental data and theoretical arguments 
(see \cite{Chomaz04} and references therein) suggest the existence of
an ``isospin fractionation'' favoring isospin symmetric IMFs, with a
corresponding neutron enrichment of LCPs and free nucleons:
indeed, also in the present data, the midvelocity emissions have
larger deuteron-to-proton and especially triton-to-proton ratios than
the evaporation. 
In any case, being the IMF multiplicities small ($\ltap$1), the
uncertainty caused by this assumption is less than one mass unit.  
Great care was also devoted to determining $A_\mathrm{sec}$ via
time-of-flight measurements with both gas and Silicon detectors:
the obtained secondary masses were always in good agreement with
the so-called Evaporation Attractor Line (EAL) \cite{Charity98}, 
except for low TKEL where the excitation is low and
the mass of the PLF$^\star$  close to the stability valley
(which is slightly more neutron-rich than the EAL).

It is relatively simple to perform a check of the energy balance of all
forward-going reaction products. 
In fact, the total energy is obtained (for each bin of TKEL) by adding
the c.m. kinetic energy of the PLF residue to the energy associated to
the emission of all forward-going particles, without distinction of
their production mechanism, namely
\begin{equation}
\langle E^\mathrm{\;c.m.}_\mathrm{\,forw}\rangle =
      \sideset{}{_i} \sum 
         \overline{k^\mathrm{\,c.m.}_\mathrm{\,C_i}}
             \langle\mathcal{M}_{\,C_i}\rangle 
     +   \overline{k^\mathrm{\,c.m.}_\mathrm{\,n}}
             \langle\mathcal{M}_{n}\rangle  \;
      - \; Q_\mathrm{\;tot}
\label{eq:ebil}
\end{equation}
where $\overline{k^\mathrm{\,c.m.}_\mathrm{\,C_i}}$
is the average (efficiency corrected) c.m. kinetic energy of the
charged particles of the $i$-th species and the sum extends over the
different species;  
$\overline{k^\mathrm{\,c.m.}_\mathrm{\,n}}$ is the average
c.m. kinetic energy of the free neutrons (estimated from the protons
after correcting for the lack of Coulomb repulsion);
$Q_\mathrm{tot}$ is the Q-value for disassembling the projectile into
the average secondary PLF, plus as many neutrons, LCPs and IMFs as
given by their respective multiplicities. 
We expect to come close to the c.m. kinetic energy of the incoming
projectile (about 885 MeV), having disregarded only the energy of
$\gamma$-rays, which are mainly emitted at the end of the evaporation
chain.  
Indeed the sum falls short of the projectile energy by less than 
15 MeV at small TKEL values (rising to about 50 MeV at
TKEL$\approx$500 MeV), thus giving confidence in the results of the
present analysis. 

%
\begin{figure}
\includegraphics[width=75mm,bb=35 20 345 205]{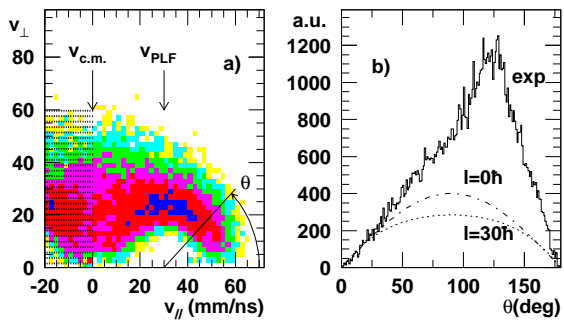}
\caption{\label{fig:theta}
Left: experimental yield of $\alpha$-particles, at TKEL=600 MeV,
in the ($v_{\,\perp}$,$v_{\,\parallel}$) plane with respect to the
PLF$^*$-TLF$^*$ separation axis.
Right: corresponding angular distribution in the PLF frame
(histogram) and results of simulations for an evaporating source with
spin 0$\hbar$ and 30$\hbar$ (dashed and dotted lines, respectively,
normalized in the range $\theta\leq 30^\circ$).
}
\end{figure}
It is much more difficult to separate the energy associated to the
midvelocity emissions from that associated to the sequential
evaporation from the fully equilibrated PLF$^*$.
First it is necessary to estimate the yields of the two components.
In peripheral collisions, this is feasible because 
the PLF can always be safely distinguished from the TLF (on the
basis of the phase-space distributions of the heavy remnants) and 
the most forward part of the LCP and IMF emission can be
ascribed to a pure evaporative process. 
These same considerations advise against extending the study to more
central collisions (that is, in the present work, beyond
TKEL$\approx$600 MeV). 
The procedure is based on the distributions of the
emission angle $\theta$ in the PLF frame
(see {\em e.g.} Fig. \ref{fig:theta}a),
taking the PLF$^*$-TLF$^*$ separation axis as polar axis.
For pure evaporation and neglecting recoil effects (which are
important only at small TKEL), one expects approximately a 
$\sin(\theta)$-distribution for a source with zero spin and a somewhat
flatter shape for non-zero spin 
(see {\em e.g.} Fig.~\ref{fig:theta}b). 
In reality, normalizing the distributions below $30^\circ$, one finds
that for all particle species the data have a large excess at backward
angles (with a tail extending well below $\theta=90^\circ$), which is
ascribed to midvelocity emissions. 
Therefore the evaporative multiplicities were extrapolated from the
experimental data measured in the range $\theta\leq 30^\circ$ 
(with the spin of the PLF$^*$ deduced from the out-of-plane 
angular distributions \cite{Mangiarotti03}).
The multiplicities associated to the midvelocity emissions
were then determined from the difference between the total
multiplicities and the evaporative ones.

It is now necessary to make also a hypothesis on the subdivision of
$\langle\mathcal{M}_{n}\rangle$ 
between the midvelocity emission,
$\langle\mathcal{M}^\mathrm{midv}_{n}\rangle$,
and the subsequent statistical evaporation of the PLF$^\star$,
$\langle\mathcal{M}^\mathrm{evap}_{n}\rangle$.
No firm conclusion has been reached in the literature concerning a
possible enhancement of neutrons in the neck matter,
however there are indications~\cite{Sobotka00} that globally the 
N/Z ratio of the neck matter is the same as that of the bulk matter.
Therefore, as a working hypothesis, the overall neck emissions
(and consequently also the primary PLF$^\star$) 
have been assumed to have the same N/Z (=2.27) of the system.

\begin{figure}
 \includegraphics[width=85mm,bb=5 20 460 375]{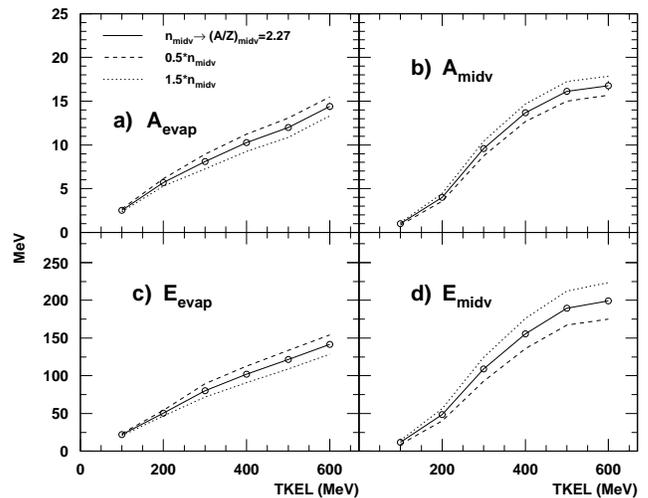}
\caption{\label{fig:bal} 
The figure presents, 
as a function of TKEL,
the average total mass of 
 (a) the PLF$^\star$ evaporation,
     $\langle A_\mathrm{\,evap} \rangle$,  and
 (b) the forward-going midvelocity emissions,
     $\langle A_\mathrm{\,midv} \rangle$;
the average amount of energy involved in
 (c) the PLF$^\star$ evaporation,
     $\langle E_\mathrm{\;evap} \rangle$, and
 (d) the forward-going midvelocity emissions,
     $\langle E_\mathrm{midv}\rangle$.
The different curves correspond to different 
repartitions of the free neutrons (see text).
}
\end{figure}

Finally, one needs the average kinetic energy for each
particle species. 
For evaporated particles this is best done in the frame of the 
emitting PLF, as the value obtained for  $\theta\leq 30^\circ$
can be used for the whole distribution.

It is now possible to estimate the total mass evaporated from the
excited PLF$^\star$ and the associated energy (measuring the initial
excitation energy) by summing up masses and kinetic energies 
(the superscript ``PLF'' reminds that they are not 
in the c.m. frame) of neutrons, LCPs and IMFs weighted with their
evaporative multiplicities 
\begin{align}
  \langle{A}_\mathrm{\,evap}\rangle &=
      \sideset{}{_i} \sum A_{C_i}
             \langle\mathcal{M}^\mathrm{\,evap}_{C_i}\rangle
     \; \; \; \; \; \; + \langle\mathcal{M}^\mathrm{\,evap}_{n}\rangle
  \label{eq:astat}
  \\
  \langle E_\mathrm{\;evap} \rangle &=
      \sideset{}{_i} \sum 
      \overline{k^\mathrm{\,PLF}_\mathrm{C_{i},evap}}
             \langle\mathcal{M}^\mathrm{evap}_{C_i}\rangle 
    + \overline{k^\mathrm{\,PLF}_\mathrm{n,evap}}
             \langle\mathcal{M}^\mathrm{evap}_{n}\rangle  \notag \\
       &\; \; \;  - \; Q_\mathrm{\;evap} 
  \label{eq:estat}
\end{align}
where $\overline{k^\mathrm{\,PLF}_\mathrm{C_{i},evap}}$
is the average (efficiency corrected) kinetic energy of the evaporated
charged particles of the $i$-th species (in the PLF frame, evaluated
for $\theta\leq30^\circ$); 
$\overline{k^\mathrm{\,PLF}_\mathrm{n,evap}}$ is the average kinetic
energy of the evaporated neutrons (estimated from that of the protons 
minus an average Coulomb barrier);
$Q_\mathrm{evap}$ is the Q-value for disassembling the average primary
PLF$^\star$ into the average secondary PLF, plus as many neutrons,
LCPs and IMFs as given by their respective evaporative multiplicities.
The obtained results are shown in Fig.~\ref{fig:bal}(a) and (c).
Here, as in the other panels, the symbols correspond to the adopted
subdivision of the free neutrons, while the dashed and dotted lines
correspond to a $\pm$50\% variation of
$\langle\mathcal{M}^\mathrm{midv}_{n}\rangle$:
the results are not very sensitive to this assumption. 
The behavior of both $\langle{A}_\mathrm{\,evap}\rangle$ and 
$\langle E_\mathrm{\;evap} \rangle$ is very regular,
displaying a steady, almost linear increase with increasing $TKEL$. 
This suggests that TKEL, even in the presence of a sizable midvelocity 
emission, remains nevertheless a good indicator 
of the average excitation energy of the PLF$^\star$ 
(and of the TLF$^\star$ as well) \cite{Mangiarotti03}.
The obtained average energetic cost per evaporated nucleon is about
9--10 MeV, in reasonable agreement with results of the statistical
code {\sc Gemini}~\cite{Charity88b}.

By subtracting from the mass and energy of all emitted particles 
the contribution of the evaporation, one can estimate the part 
pertaining to the midvelocity emissions
\begin{align}
    \langle{A}_\mathrm{\,midv}\rangle &= 
              \left( \sideset{}{_i} \sum A_{C_i}
                     \langle\mathcal{M}_{C_i}\rangle
                   + \langle\mathcal{M}_{n}\rangle   \right)
            - \langle{A}_\mathrm{\,evap}\rangle
   \label{eq:amidv}
   \\
   \langle E_\mathrm{midv}\rangle &=
        \langle E^\mathrm{\;c.m.}_\mathrm{forw}\rangle -
        \left( \langle E_\mathrm{\;evap}   \rangle + 
               \langle K_\mathrm{\;transl} \rangle \right)
    \label{eq:emidv}
\end{align}
As the kinetic energies of 
$\langle E^\mathrm{\;c.m.}_\mathrm{forw}\rangle$
are evaluated in the c.m. reference frame and those of 
$\langle E_\mathrm{\;evap} \rangle$ in the PLF frame, the
translational kinetic energy of the whole pattern of evaporated
particles (due to the motion of the source) was taken into account
with the term
 $\langle K_\mathrm{\;transl} \rangle \approx
     \frac{1}{2}\; m_N \; \langle A_\mathrm{\,evap} \rangle \; 
      \langle v^{\,2} \rangle$,
where $m_N$ is the nucleon mass, $A_\mathrm{\,evap}$ the mass
number of the total evaporation from PLF$^\star$ and
$v$ the c.m. velocity of the PLF residue
\footnote{ This neglects corrections 
        (\protect{$\ltap$}10 MeV) due to correlations
        of \protect{$A_\mathrm{evap}$} with \protect{$v^{2}$};
        the correct expression  
        \protect{$\langle K_\mathrm{transl}\rangle=\frac{1}{2} 
         m_N  \langle  A_\mathrm{evap}\; v^{2} \rangle$},
        is actually unusable even with 4$\pi$ detectors,
        as it requires to identify the origin of each particle.
        }.

The results of Eq.~\ref{eq:amidv} and \ref{eq:emidv} 
are shown in Fig.~\ref{fig:bal}(b) and (d), respectively. 
With increasing TKEL, both $\langle{A}_\mathrm{\,midv}\rangle$ and 
$\langle E_\mathrm{midv}\rangle$ first increase almost linearly and
then flatten (at values around 16 and 180 MeV, respectively).
The origin of this flattening is still an open question and will be
the subject of further investigation 
\footnote{The flattening of 
   \protect{$\langle{A}_\mathrm{\,midv}\rangle$} may be 
    partly induced by the selection of binary events: 
    events with a more massive interaction zone are likely 
    to emit larger fragments and have a larger chance of
    being classified as 3-body events.}. 
The comparison of Fig.~\ref{fig:bal}(c) and (d), shows that, at a
given TKEL, up to about half of the energy goes into the midvelocity
component, which is really an essential aspect of the reaction and may
represent \cite{Baran04} the prodrome of the multifragmentation of the
whole system, occurring in central collisions.

In short, it can be stated that $\langle E_\mathrm{\,midv}\rangle$ and
$\langle E_\mathrm{\,evap} \rangle$ are comparable and this statement
holds in spite of several systematic uncertainties that affect their
evaluation (conservatively, altogether up to 30\% of the quoted values). 
So in the Fermi domain, an important part of the dissipated energy,
initially stored in the translational motion of the projectile, is
localized in the new midvelocity emission. 
This mechanism has an important role in the overall balance of the
reaction, both in terms of the emitted mass (charge) and energy
\cite{Gobbi}. 

Finally, since the amount of dissipated energy localized in the
midvelocity matter is comparable to that in the PLF$^\star$ or 
TLF$^\star$, and the mass of the emitting zone 
(two sources as schematized in Ref. \cite{Piantelli02} or 
--more realistically-- a whole distribution of sources)
is certainly smaller, one may expect for the midvelocity matter a
value of $E^\star/A$  
(of the order of 7--14 MeV, depending on the assumed source size) 
significantly larger than that of the evaporative source 
($\ltap$2 MeV for the data of this work). 
It is well known experimentally that with increasing $E^\star/A$
the disassembly properties of nuclear matter radically change, 
leading to a preferential formation of IMFs~\cite{Ogilvie91,Beaulieu96}.
A large $E^\star/A$ value is therefore consistent with
the observation of a preferential emission of IMFs at midvelocity. 
Experimental data on the deposition of energy in the midvelocity
matter may represent an important benchmark for the most sophisticated
microscopic calculations \cite{Feldmeier90,Ono92,Ono99,Feldmeier00}.
When such comparisons with the data become available, 
one may hope to gain some new insight on the mechanism 
at the basis of the large, strongly localized, energy deposition 
as well as on the transport of internal energy from the
interaction zone to the cold projectile and target remnants.



\begin{acknowledgments}
%
Many thanks are due to the staff of LNS for continuous support and for
providing very good pulsed beams.
\end{acknowledgments}

\bibliography{letter}
\end{document}